\documentclass[aps,prd,preprint,superscriptaddress,
tightenlines,nofootinbib]{revtex4}

\usepackage{graphicx}
\usepackage{dcolumn}
\usepackage{bm}
\usepackage{epsfig}

\def \numups{(5.81\pm0.12)\times 10^{6}}

\def \effone{6.81}
\def \erreffone{\pm 0.07}
\def \efftwo{6.23}
\def \errefftwo{\pm 0.06}

\def \pyieldone{32.6}
\def \pyieldtwo{20.1}
\def \peyo{\asy{6.9}{6.1}}
\def \peyt{\asy{5.8}{5.1}}

\def \rat{0.62}
\def \erat{\asy{0.27}{0.20}}

\def \corrat{0.67}
\def \ecorrat{\asy{0.30}{0.22}}
\def \tot{52.4}
\def \etot{\asy{7.5}{6.9}}

\def \prodone{0.82}
\def \stprodone{\asy{0.17}{0.15}}
\def \sysprodone{\pm 0.06}

\def \prodtwo{0.55}
\def \stprodtwo{\asy{0.16}{0.14}}
\def \sysprodtwo{\pm 0.04}

\def \brone{1.63}
\def \stbrone{\asy{0.35}{0.31}}
\def \sysbrone{\asy{0.16}{0.15}}

\def \brtwo{1.10}
\def \stbrtwo{\asy{0.32}{0.28}}
\def \sysbrtwo{\asy{0.11}{0.10}}

\def \prd#1#2#3{{Phys. Rev. D} {\bf#1}, #2 (#3)}
\def \prl#1#2#3{{Phys. Rev. Lett.} {\bf#1}, #2 (#3)}
\def \plb#1#2#3{{Phys. Lett. B} {\bf #1} #2 (#3)}
\def \nima#1#2#3{{\it Nucl. Instr. Meth.} {\bf A#1} #2 (#3)}
\def \zpc#1#2#3{{\it Zeit. Phys.} C {\bf#1} (#3) #2}
\def \etal{{\it et\,al.}}

\def \asy#1#2{{^{+#1}_{-#2}}}
\newcommand{\plus}{\mbox{$^{+}$}}
\newcommand{\minus}{\mbox{$^{-}$}}

\newcommand{\goesto}{\mbox{$\rightarrow$}}

\newcommand{\lplus}{\mbox{$\ell^+$}}
\newcommand{\lminus}{\mbox{$\ell^-$}}
\newcommand{\dilep}{\mbox{$\lplus\lminus$}}
\newcommand{\bbar}{\mbox{$\bar{b}$} }
\newcommand{\bbbar}{\mbox{$b\bbar$}}  
\newcommand{\upsot}{\mbox{$\Upsilon$}{\rm (1S,2S)}}
\newcommand{\upsi}{\mbox{$\Upsilon$}{\rm (1S)}}
\newcommand{\upsii}{\mbox{$\Upsilon$}{\rm (2S)}}
\newcommand{\upsiii}{\mbox{$\Upsilon$}{\rm (3S)}}
\newcommand{\upsns}{\mbox{$\Upsilon$}{\rm (nS)}}

\newcommand{\etab}{\mbox{$\eta_b$}}
\newcommand{\chib}{\mbox{$\chi_{b}(1P)$}}
\newcommand{\chibp}{\mbox{$\chi_{b}(2P)$}}

\newcommand{\chibpzero}{\mbox{$\chi_{b0}(2P)$}}
\newcommand{\chibpone}{\mbox{$\chi_{b1}(2P)$}}
\newcommand{\chibptwo}{\mbox{$\chi_{b2}(2P)$}}
\newcommand{\chibpot}{\mbox{$\chi_{b1,2}(2P)$}}
\newcommand{\chibpj}{\mbox{$\chi_{bJ}(2P)$}}
\newcommand{\piz}{\mbox{$\pi$}^{0}}

\newcommand{\pipi}{\mbox{$\pi$}^{+}\mbox{$\pi$}^{-}}
\newcommand{\pizpiz}{\mbox{$\pi$}^{0}\mbox{$\pi$}^{0}}
\newcommand{\mev}{\mbox{ MeV}}
\newcommand{\gev}{\mbox{ GeV}}

\newcommand{\gevc}{\mbox{ GeV/c}}

\newcommand{\epm}{\mbox{$e^\pm$}}

\newcommand{\jpc}{\mbox{$J^{PC}$}} 
\newcommand{\omm}{\mbox{$1^{--}$}}


\begin{document}
\preprint{CLNS 03/1840}   
\preprint{CLEO 03-12}   

\title{
Observation of the Hadronic Transitions
$\chibpot\goesto\omega\upsi$
}

\author{D.~Cronin-Hennessy}
\author{C.~S.~Park}
\author{W.~Park}
\author{J.~B.~Thayer}
\author{E.~H.~Thorndike}
\affiliation{University of Rochester, Rochester, New York 14627}
\author{T.~E.~Coan}
\author{Y.~S.~Gao}
\author{F.~Liu}
\author{R.~Stroynowski}
\affiliation{Southern Methodist University, Dallas, Texas 75275}
\author{M.~Artuso}
\author{C.~Boulahouache}
\author{S.~Blusk}
\author{E.~Dambasuren}
\author{O.~Dorjkhaidav}
\author{R.~Mountain}
\author{H.~Muramatsu}
\author{R.~Nandakumar}
\author{T.~Skwarnicki}
\author{S.~Stone}
\author{J.C.~Wang}
\affiliation{Syracuse University, Syracuse, New York 13244}
% ...
\collaboration{CLEO Collaboration} 
\noaffiliation


\date{\today}

\begin{abstract} 
The CLEO Collaboration has made the first observations of 
hadronic 
transitions among bottomonium ($\bbbar$) states
other than the dipion transitions among $\upsns$ states.
In our study of $\upsiii$ decays, 
we find a significant signal for 
$\upsiii\goesto\gamma\omega\upsi$ 
that is consistent with radiative decays $\upsiii\goesto\gamma\chibpot$,
followed by $\chibpot\goesto\omega\upsi$.
The branching ratios we obtain are 
${\cal{B}}(\chibpone\goesto\omega\upsi)$
$=(\brone\stbrone\sysbrone)\%$
and
${\cal{B}}(\chibptwo\goesto\omega\upsi)$
$=(\brtwo\stbrtwo\sysbrtwo)\%$,
in which the first error is statistical and  
the second is systematic.
\end{abstract}

\maketitle

\begin{figure}
\includegraphics[width=\linewidth]{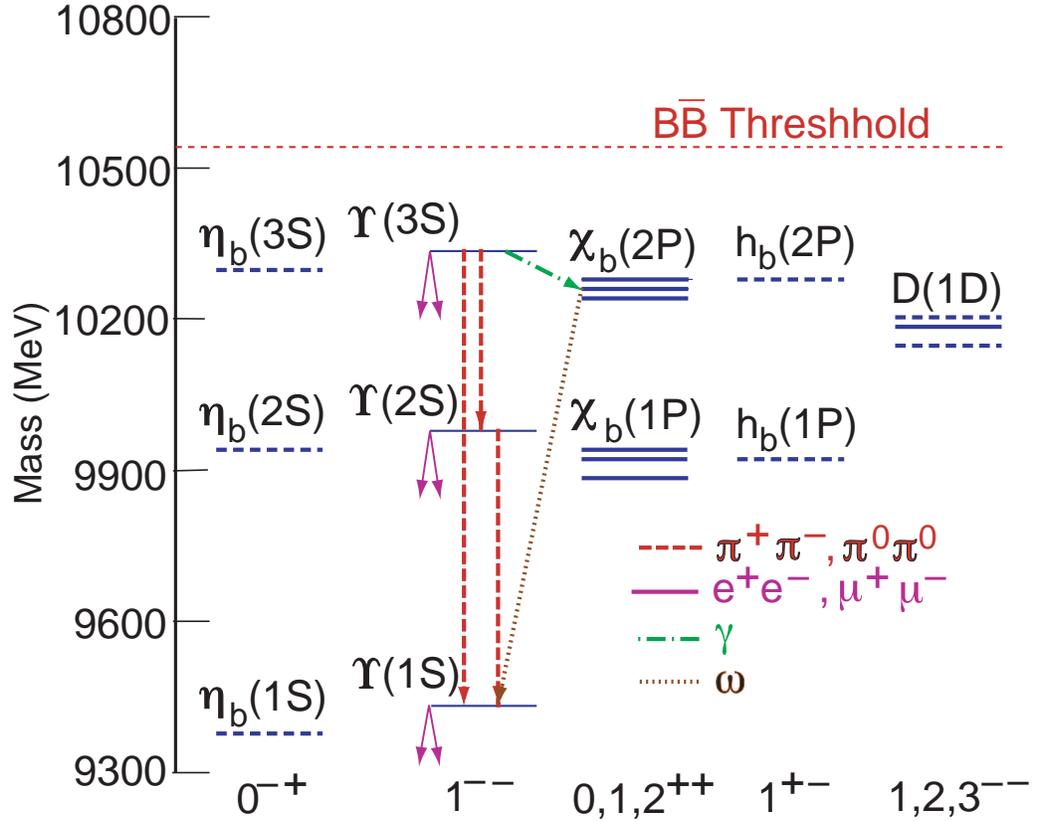}
\caption{The spectrum of bottomonium states below open bottom
threshold. Dilepton decays of 
the $\jpc=\omm$ states are denoted by the solid lines, while dipion
transitions are denoted by dashed lines. The radiative decay
$\upsiii\goesto\gamma\chibpot$ is indicated by the dot-dashed line, and
the decay $\chibpot\goesto\omega\upsi$ by
the dotted line.\label{bbspec}}
\end{figure}

The only hadronic decays of bottomonia that have
been experimentally observed to date are the 
$\pi\pi$ ($\pi\pi\equiv\pipi$ and $\piz\piz$) transitions among
the $\upsns$ states~\cite{expt3s}.
In Fig. ~\ref{bbspec} we show the spectrum of 
bottomonium states below open bottom threshold.

Hadronic transitions among heavy quarkonia are 
generally understood to proceed by the emission of 
soft gluons, and
subsequent hadronization of the gluons.  
The analysis of heavy quarkonium hadronic transitions 
is one of few possible laboratories for the study of 
the physics of the soft gluon emission and 
hadronization process that governs such decays. 

Most theoretical work dedicated to these transitions has been built around
a multipole expansion of the color field, an idea first proposed
by Gottfried and Yan~\cite{gottyan}.
A fairly substantial literature exists which attempts to 
describe in detail the $\pi\pi$ transitions that have been
observed~\cite{theory}.
While these transitions do provide
important information about strong interaction 
dynamics in heavy quark systems, the investigation
of other hadronic decay modes ({\em i.e.}, involving $\eta$, 
$\omega$ or multiple $\pi$) should offer a different perspective.

In the multipole expansion model,
a hadronic transition involving $\omega$ requires
three gluons in an $E1* E1* E1$ 
configuration~\cite{yanpriv}.  For such a purely electric coupling, 
Voloshin~\cite{volo} has recently predicted roughly equal rates for the 
decay of the two 
states $\chibpone$ and $\chibptwo$ to
$\omega\upsi$.
In this Letter, we report on the observation of 
the transitions $\chibpot\goesto\omega\upsi$.

The data set consists of 
$\numups\upsiii$ decays observed with 
the CLEO III~\cite{cleo3det} detector at the Cornell Electron Storage Ring.
Charged particle tracking is done by the 47-layer drift 
chamber and a four-layer silicon 
tracker which reside in a 1.5~T solenoidal magnetic field.
Photons are detected using an electromagnetic calorimeter
consisting of 7784 CsI(Tl) crystals distributed in a 
projective barrel geometry.
The particle-identification capabilities of the CLEO III
detector are not used in the present analysis.


Events consistent with the 
final state of $\gamma\pipi\piz\dilep$ 
are selected by 
requiring two high momentum charged tracks ($p > 4 \gevc$) 
and two or three charged low momentum tracks
($0.12 < p < 0.75 \gevc$).
The low momentum tracks 
are required to come from the interaction region
using criteria obtained by studying charged pion tracks 
in a sample of events from the 
kinematically similar decay $\upsii\goesto\pipi\upsi$.

We require events to contain an $\upsi$ candidate
by requiring that the two high momentum tracks in the
event have an invariant mass in the range 9300 to 9600 MeV,
consistent with 
the $\upsi$ mass.
We make no cuts on the track quality variables for 
the lepton candidate tracks, nor do we
require them to satisfy lepton-identification criteria.
The invariant mass requirement alone 
provides a nearly background free sample,
and imposition of lepton-identification criteria 
only leads to larger systematic uncertainties and reduced
signal efficiency without much improvement in 
signal quality.

We require events to have three or four 
showers in the calorimeter, each of which 
has $E>30$ MeV, 
and is not matched to any 
charged track.  Two of these showers must
form an invariant mass within three standard deviations 
$(3\sigma)$ of the known
$\piz$ mass.  These candidates are kinematically constrained
to the known $\piz$ mass, in order to improve the 
momentum resolution of the $\piz$ candidates.
In addition to the two showers that correspond to the 
$\piz$, events must contain an isolated photon candidate,
between $50$ and $250$ MeV in energy, that does not 
form an invariant mass within $8 \mev$ $(1.5 \sigma)$
of the $\piz$ mass
with any other shower.  Furthermore we require that 
the polar angle $(\theta)$ of the third shower satisfy 
$|\cos{\theta}| < 0.804$, 
the angular region in which CLEO's energy resolution is best.
We allow events to contain up to one additional shower 
in the range $|\cos{\theta}|<0.804$.  
In addition, we allow for the possibility of one spurious charged 
track candidate in addition to the four ``signal'' tracks.
Such spurious tracks may arise from failures in pattern recognition
or from delta rays.  Spurious showers may arise from synchrotron 
radiation from the $\epm$ beams or as a result of 
random noise in the calorimeter.
If a given event yields more than one candidate due to
the presence of an additional shower or track,
we choose the candidate for which the sum of energies of
all final state particles is nearest $M(\upsiii)$.

\begin{figure}[t]
\begin{center}
\includegraphics[width=\linewidth]{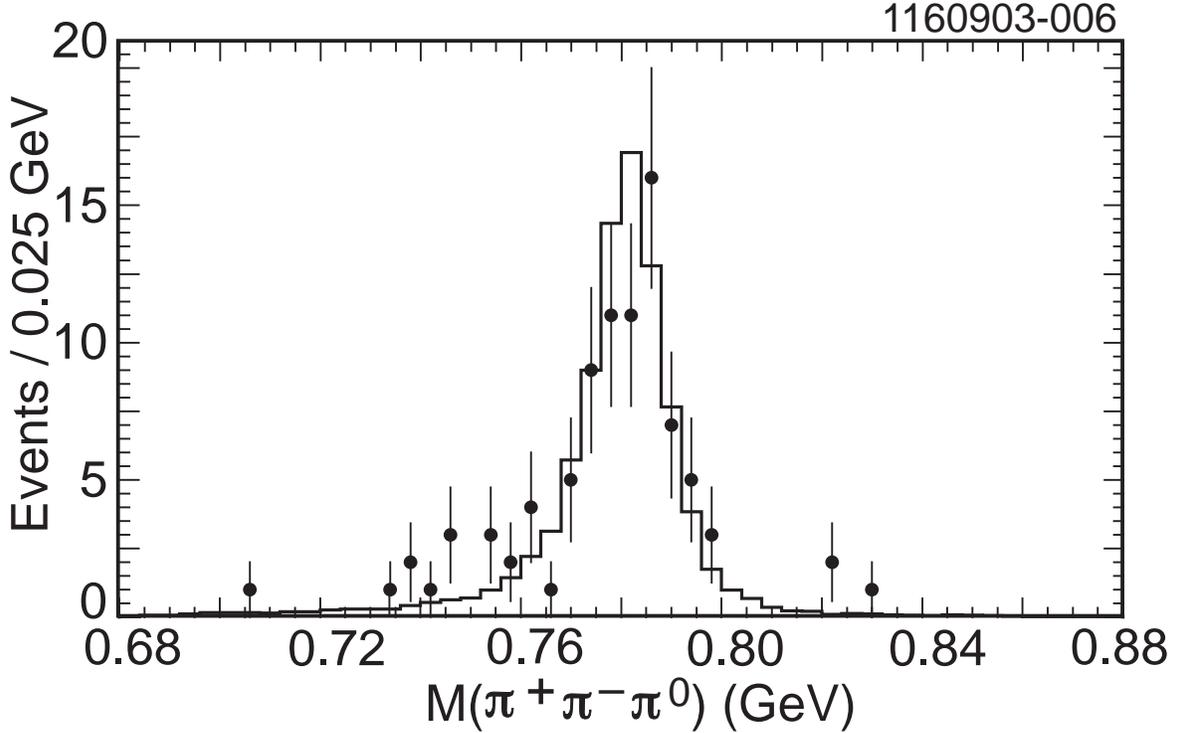}
\end{center}
\caption{The $\pipi\piz$ invariant mass, for data events 
subject to final analysis cuts with the exception of 
the cut on the $\chi^2/d.o.f.$ 
of the kinematic fit to $\omega$. \label{3pimass}
The overlaid histogram shows signal Monte Carlo events
(normalized to the same total number) and indicates the 
good reproduction by the Monte Carlo of the 
shape and location of the $\omega$ peak.}
\end{figure}

Because there is no phase space
for a pair of kaons for decays in which a 
$\upsi$ is present,  we assume that the 
low momentum charged tracks are pions.
The invariant mass of the $\pipi\piz$ combination,
plotted in Fig.~\ref{3pimass},
exhibits a clear 
enhancement at the mass of $\omega$, 
$M(\omega)=0.783 \gev$~\cite{pdg2002}.

To complete the reconstruction of the full
decay chain,
$\upsiii\goesto\gamma\omega\upsi\goesto
\gamma\pipi\piz\dilep$, 
we require that the $\chi^2/d.o.f.$ be less than two 
for a kinematic fit of $\pipi\piz$ constrained 
to the $\omega$ mass,
and subsequently that the
mass recoiling against the
kinematically fitted $\omega$ candidate 
and the photon
lies within $\asy{25}{20}$ MeV of $M(\upsi)=9.460 \gev$~\cite{pdg2002}.  

The simplest explanation for the observed events is the
decay sequence $\upsiii\goesto\gamma\chibpot$, with 
$\chibpot\goesto\omega\upsi$. 
The lowest mass $\chibpj$ state, 
$\chibpzero$, lies below threshold for decay to $\omega\upsi$.
In principle a transition through the $\etab(3S)$ state
(see Fig.~\ref{bbspec}) is possible, but this state has
never been observed, and, furthermore, the energy of the 
photon in the decay $\upsiii\goesto\gamma\etab(3S)$ is 
expected to be below the range of observed energies in 
the data.  


Backgrounds from ordinary $udsc$ quark pair production are 
extremely small because of the presence 
of the $\upsi\goesto\dilep$ decay 
in the signal sample.  The only significant source of 
background $\upsi$ expected is known cascades from $\upsiii$.
The final state of $\gamma\pipi\piz\upsi$ may be reached through:
\begin{eqnarray*}
\upsiii\goesto\gamma\chibp,\;
\chibp\goesto\gamma\upsii,\;
\upsii\goesto\pipi\upsi\\
\mbox{or}\;\;\upsiii\goesto\piz\piz\upsii,\;
\upsii\goesto\pipi\upsi.
\end{eqnarray*}
In the first case, the final state of interest 
can be produced by 
the addition of a spurious shower in the calorimeter.
In the second case, it 
may be reached by loss of one photon from one of the neutral pions
due to acceptance or energy threshold.

Backgrounds produced through either of these 
two processes are removed by excluding events in which
the mass recoiling against the two charged pions 
in the $\upsiii$ reference frame 
is consistent ({\em i.e.}, between 9.78 and 9.81 GeV) 
with the hypothesis that the 
$\pipi$ system is recoiling against 
$\upsii$ in the process $\upsiii\goesto X+\upsii
\goesto X+\pipi\upsi$.

Two other decay sequences can yield
the final state of 
$\gamma\pipi\piz\upsi$:
the decay $\upsiii\goesto\pipi\upsii$, with
the $\upsii$ decaying either to $\pizpiz\upsi$,
or to $\gamma\chib$ followed by $\chib\goesto\gamma\upsi$.
In each of these cases, however, the charged pions are
too soft to produce spurious $\omega$ candidates
for the signal decay chain.

In order to evaluate residual background, we 
generated a GEANT~\cite{geant} Monte Carlo sample for the channel
$\upsiii\goesto\gamma\chibpot$,$\chibpot\goesto\gamma\upsii$,
$\upsii\goesto\pipi\upsi$,
corresponding to $21.5 \pm 3.1$ million $\upsiii$ decays,
or $4.53\pm0.65$ times our data set.
The uncertainty on the equivalent number of 
$\upsiii$ decays is due to the error on the branching
ratios needed to convert our number of generated events
to the equivalent number of $\upsiii$ decays.
This Monte Carlo sample produced one event
that satisfied our selection criteria.
We therefore expect $0.22 \pm 0.03$ events 
due to this source.  We also generated a Monte Carlo sample of
$\upsiii\goesto\piz\piz\upsii,\;\upsii\goesto\pipi\upsi$,
corresponding to $540\asy{110}{80}$ million $\upsiii$ decays,
or $113\asy{23}{16}$ times our data set.
From this sample, a total of nine events passed our selection.
In our data set, we thus 
expect $0.08 \pm 0.01$ events 
due to this source.
To account for the residual background, 
we subtract the expected contribution of
0.30 events from the observed yield. 
We conservatively set a systematic error 
of $\pm 0.30$ events due to this subtraction.

To evaluate the signal detection efficiency $\epsilon$, we 
generated 150,000 Monte Carlo events 
for each of $\chibpone$ and
$\chibptwo$, proceeding through the sequence
$\upsiii\goesto\gamma\chibpot\goesto
\gamma\omega\upsi\goesto\gamma\pipi\piz\dilep$,
and uniform angular 
distributions for the $\upsiii\goesto\gamma\chibpot$ 
and $\upsi\goesto\dilep$ decays. The masses for all
particles in the decay chain 
were taken from Ref.~\cite{pdg2002}.

The analysis cuts described above are 
applied to these samples, and we obtain
$\epsilon(\chibpone) = (\effone\erreffone)\%$
and
$\epsilon(\chibptwo) = (\efftwo\errefftwo)\%$,
including all selection criteria, acceptance and trigger efficiencies.
We apply an additional relative systematic error of $\asy{0}{3}\%$
to the efficiency in order to account for the possibility that
the $\upsi$ retains the initial polarization of the $\upsiii$.   

\begin{figure}
\includegraphics[width=\linewidth]{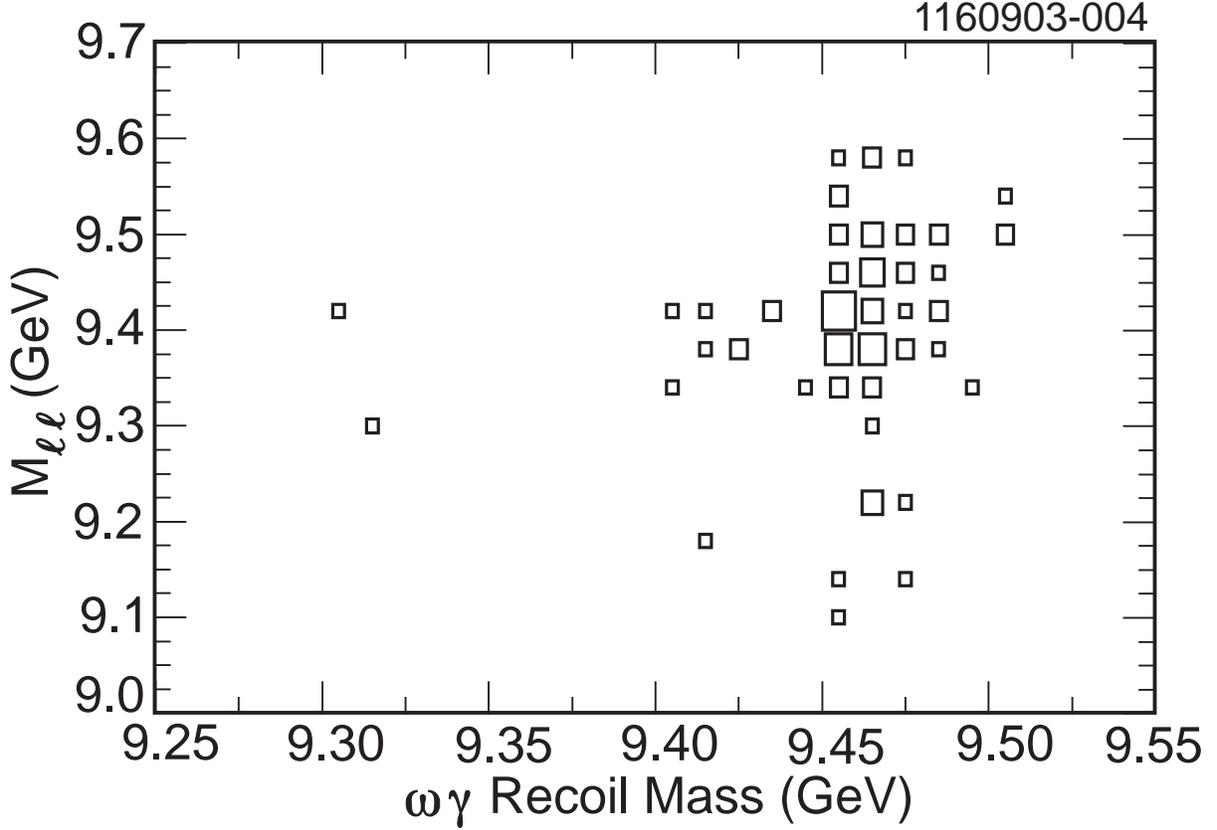}
\caption{Plot of recoil mass versus dilepton mass, for data events
subject to the final set of cuts, with the exception of 
the cuts on the two variables plotted.  
The number of events represented by each square range from one to four.
~\label{boxplot}}
\end{figure}

\begin{figure}
\includegraphics[width=\linewidth]{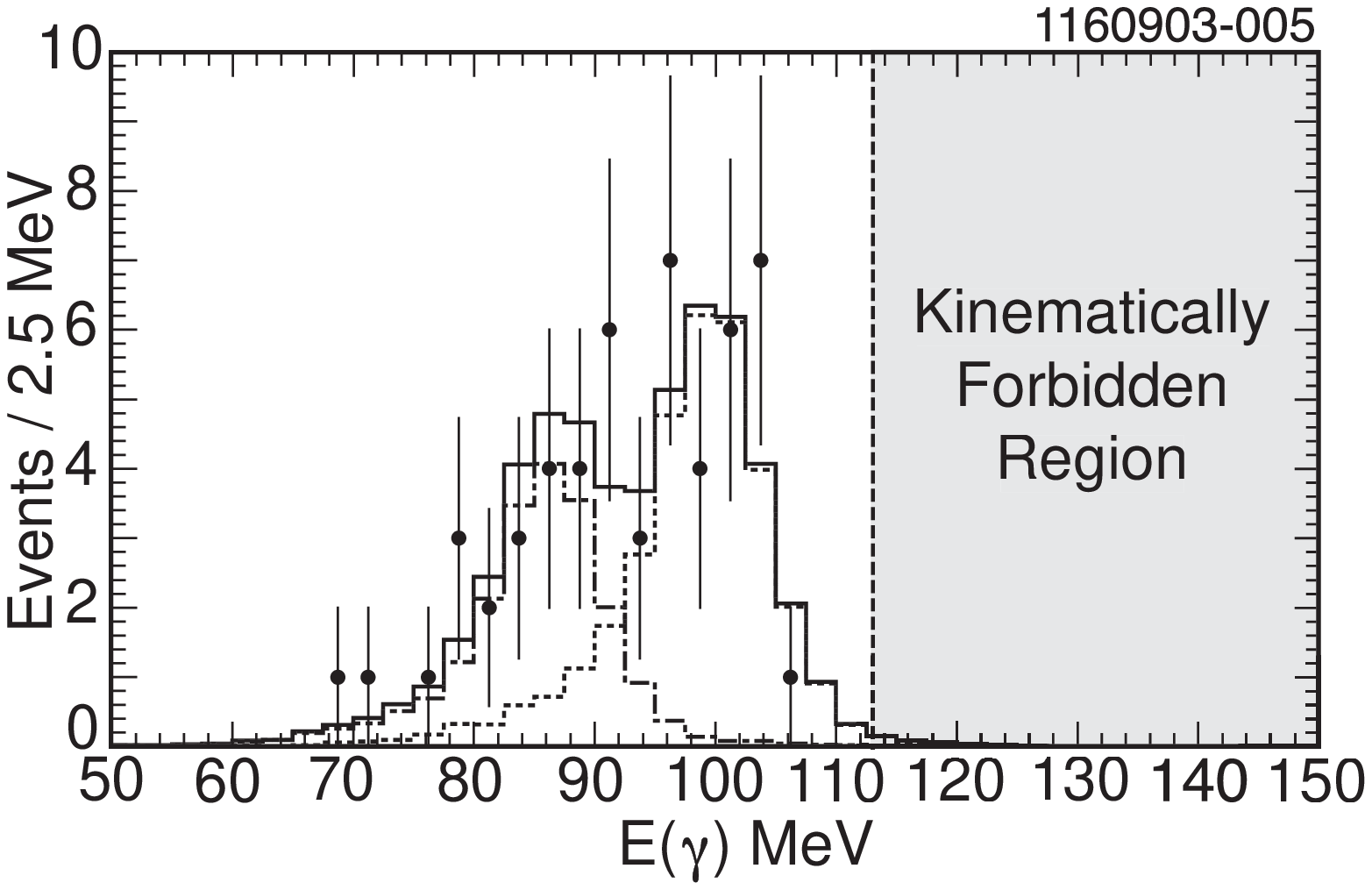}
\caption{Fitted photon energy spectrum for the
final selection of events.  The solid histogram shows contributions
for both $\chibpone$ and $\chibptwo$, while the 
dotted and dashed histograms show the individual $\chibpone$ and
$\chibptwo$ contributions, repsectively.
The dotted line indicates the region above which $\gamma$ 
energies are disallowed for lack of phase space.  
~\label{fitboth}}
\end{figure}

In order to illustrate the purity of the signal, 
in Fig.~\ref{boxplot}, we present a scatter plot 
of the mass recoiling against the $\omega\gamma$ system versus the 
dilepton invariant
mass for all events subject to all the cuts discussed above,
except those on the variables plotted.
The final $E_{\gamma}$ spectrum is
shown in Fig. ~\ref{fitboth}.
The observed yield has possible contributions from both
decay sequences involving $\chibpone$ and $\chibptwo$ intermediate
states. 

To obtain branching fractions for the 
$\chibptwo$ and $\chibpone$ transitions, 
we perform a maximum likelihood fit of
the $E_{\gamma}$ spectrum.
The expected photon spectra for $\upsiii$ transitions to 
$\chibptwo$ and $\chibpone$ were obtained from 
the signal Monte Carlo samples, and the observed
$E_{\gamma}$ spectrum was then fit to normalized Monte Carlo 
lineshapes with intensities (or yields) for $\chibpone$ and $\chibptwo$ 
as the free parameters. 
We obtain yields of $\pyieldone\peyo$ and $\pyieldtwo\peyt$ events,
respectively. These yields have statistical significances of 
$10.2\sigma$ and $5.2\sigma$, respectively, obtained by comparing
the likelihood of our final fitted yield to those of 
fits with zero signal events. 
The histogram resulting from the best fit is shown superimposed
on the data in Fig.~\ref{fitboth}. 


The expected background contribution of 
0.30 events is subtracted from the fitted yield
by assuming that it scales as the
ratio of the individual yields to the total yield.

We thus obtain the following product branching ratios, using the
detection efficiency and number of $\upsiii$ decays discussed above:
\begin{eqnarray}
{\cal{B}}(\upsiii\goesto\gamma\chibpone)
\times
{\cal{B}}(\chibpone\goesto\omega\upsi)\nonumber\\
\times
{\cal{B}}(\omega\goesto\pipi\piz)
\times
{\cal{B}}(\upsi\goesto\dilep)\nonumber\\
= (\prodone\stprodone\sysprodone)\times 10^{-4},\;\mbox{and}\\
{\cal{B}}(\upsiii\goesto\gamma\chibptwo)
\times
{\cal{B}}(\chibptwo\goesto\omega\upsi)\nonumber\\
\times
{\cal{B}}(\omega\goesto\pipi\piz)
\times
{\cal{B}}(\upsi\goesto\dilep)\nonumber\\
= (\prodtwo\stprodtwo\sysprodtwo)\times 10^{-4},
\end{eqnarray}
in which the first uncertainty is statistical and the second is 
systematic.

The statistical error of nearly $20\%$ is dominant.  
Systematic error contributions to the above
product of branching ratios are the following:
$2\%$ uncertainty in the number of $\upsiii$,
$1\%$ per charged track (a total of $4\%$) for track finding,
$5\%$ for $\piz$ reconstruction, $2\%$ for radiative $\gamma$
reconstruction, $1\%$ for Monte Carlo statistics,
$\asy{3}{0}\%$ for the assumption of uniform $\upsi\goesto\dilep$
angular distribution,
and $1\%$ for background subtraction.  
These contributions, added in quadrature, result in an
overall relative systematic error of 
$\asy{7.7}{7.1}\%$.

Using the present world average 
branching fractions~\cite{pdg2002} for 
$\upsiii\goesto\gamma\chibpot$,
$\upsi\goesto\dilep$ (taken to be 
twice that of 
$\upsi\goesto\mu\plus\mu\minus)$,
and $\omega\goesto\pipi\piz$,
we obtain
\begin{eqnarray}
{\cal{B}}
(\chibpone\goesto\omega\upsi)
=(\brone\stbrone\sysbrone)\%\;\mbox{     and}\\
{\cal{B}}
(\chibptwo\goesto\omega\upsi)
=(\brtwo\stbrtwo\sysbrtwo)\%,
\end{eqnarray}
The systematic errors include the additional uncertainty on 
the branching ratios for
$\upsiii\goesto\gamma\chibpot$, $\omega\goesto\pipi\piz$ 
and $\upsi\goesto\dilep$, 
which 
contribute at the level of $5.9\%$.

We may also calculate the ratio of $\chibptwo$ to
$\chibpone$ branching ratios, for which several of the
systematic errors discussed above cancel.
We obtain this through a maximum likelihood fit
to the $E_{\gamma}$ spectrum, in which the two free
parameters are the sum of yields and the ratio of
$\chibptwo$ to $\chibpone$ yields.
When this fit is performed, we obtain a sum of 
yields equal to 
$\tot\etot$ and a ratio of $\rat\erat$.
In order to convert the ratio of yields 
to the ratio of branching ratios, we 
multiply the yield ratio by 
a factor of $\epsilon(\chibpone)/\epsilon(\chibptwo)$.
\begin{eqnarray}
{\cal{B}}(\chibptwo\goesto\omega\upsi))/
{\cal{B}}(\chibpone\goesto\omega\upsi)\\
=\corrat\ecorrat.
\end{eqnarray}

The only systematic errors which do not cancel in this ratio
are the small uncertainties in 
efficiency and 
$\upsiii\goesto\gamma\chibpot$ branching
ratios.  These are entirely negligible compared to
the output statistical error from the maximum likelihood fit.  

In Ref.~\cite{volo}, Voloshin predicts on the basis of
S-wave phase space factors, for 
$E_{1}* E_{1}* E_{1}$ gluon configurations
expected by the multipole expansion model~\cite{yanpriv},
that 
$\Gamma(\chibptwo\goesto\omega\upsi)/\Gamma(\chibpone\goesto\omega\upsi)
\approx 1.4.$  
The ratio of full widths $\Gamma(\chibptwo)/\Gamma(\chibpone)$ 
lies in the range of 1.25-1.5, using world average 
measurements of ${\cal{B}}(\chibpot\goesto\gamma\upsot)$ and 
theoretical predictions for the rates
$\Gamma(\chibpot\goesto\gamma\upsot)$~\cite{predwidths}.  
Thus the branching ratios ${\cal{B}}(\chibpot\goesto\omega\upsi)$
are expected to be approximately equal.
Our measurement is in agreement with this expectation.



In summary, 
we have made the observation of the first hadronic decay modes 
involving non-vector $\bbbar$ states using a sample of 
$\numups\upsiii$ decays collected by CLEO III.
The discovery comes in the phase-space limited 
channels $\chibpot\goesto\omega\upsi$. We find that 
the ratio of the 
measured branching ratios for the two transitions 
are in agreement with 
theoretical expectations based
on S-wave phase space factors for multipole expansions. 

We gratefully acknowledge the effort of the CESR staff 
in providing us with
excellent luminosity and running conditions.
This work was supported by 
the National Science Foundation,
the U.S. Department of Energy,
the Research Corporation,
and the 
Texas Advanced Research Program.


\begin{thebibliography}{99}

\bibitem{expt3s} See, {\it e.g.}, 
F. Butler \etal, \prd{49}{40}{1994}; 
I. Brock \etal, \prd{43}{1448}{1991};
J. Green \etal, \prl{49}{617}{1982};
Q. Wu \etal, \plb{301}{307}{1993};
G. Mageras \etal, \plb{118}{453}{1982}.
\bibitem{gottyan}
K. Gottfried, \prl{40}{598}{1978};
T. M. Yan, \prd{22}{1652}{1980}.
\bibitem{theory}
Y. P. Kuang and T. M. Yan, \prd{24}{2874}{1981};
M. Voloshin and V. Zakharov, \prl{45}{688}{1980};
V. A. Novikov and M. A. Shifman 
\zpc{8}{43}{1981};
M. E. Peskin, in 
``Dynamics and Spectroscopy at High Energy'', 
Proceedings of the Eleventh SLAC Summer Institute on Particle Physics, 
SLAC Report 267, 1984;
P. Moxhay, \prd{39}{3497}{1989};
H-Y. Zhou and Y.-P. Kuang, \prd{44}{756}{1991};
G. B\'{e}langer, T. DeGrand and P. Moxhay, \prd{39}{257}{1989};
S. Chakravarty and P. Ko, \prd{48}{1205}{1993};
S. Chakravarty, S. M. Kim and P. Ko, \prd{50}{389}{1994};
T. Komada, M. Ishida and S. Ishida, \plb{508}{31}{2001}.
\bibitem{yanpriv} T-M. Yan, {\em private communication}, 2003.
\bibitem{volo} M. Voloshin, {\em Mod. Phys. Lett}{\b A} 18,1067 (2003).
\bibitem{cleo3det} CLEO Collaboration, CLNS-94-1277; D. Peterson \etal, 
\nima{478}{142}{2002}.
\bibitem{pdg2002} Particle Data Group, K. Hagiwara \etal, \prd{66}{1}{2002}.
\bibitem{geant} R. Brun \etal, GEANT 3.15, CERN Report No. DD/EE/84-1 (1987). 
\bibitem{predwidths} Obtained by using the radiative branching fractions 
of $\chibpot$ from Ref.~\cite{pdg2002} and predictions for their E1 partial 
widths found in W. Kwong and J. L. Rosner, \prd{38}{279}{1988} and
S. Gupta, S. F. Radford and W. W. Repko, \prd{34}{201}{1986}. 
\end{thebibliography}
\end{document}